\documentclass[amsmath,12pt]{article}

\begin{document}

\begin{center}

{\Large \bf
A relativistic action-at-a-distance description of gravitational interactions?}
\vskip .6in

Domingo J. Louis-Martinez
\vskip .2in

Department of Physics and Astronomy,\\ University of British
Columbia\\Vancouver, Canada, V6T 1Z1 

martinez@phas.ubc.ca

\end{center}
 
\vskip 1cm
 
\begin{abstract}
It is shown that certain aspects of gravitation may be described using a relativistic action-at-a-distance formulation.
The equations of motion of the model presented are invariant under Lorentz transformations and agree with the equations
of Einstein's theory of General Relativity, at the first Post-Newtonian approximation, for any number of 
interacting point masses. 
\end{abstract}


{\it Introduction}

After the discovery of the action-at-a-distance formulation of
electrodynamics\cite{schw} - \cite{barut}, several relativistic non-instantaneous 
action-at-a-distance
theories have been investigated \cite{dettman} - \cite{strings}. Instantaneous
action-at-a-distance formulations have been studied using a variety
of approaches \cite{hill} - \cite{droz}. For gravity, several relativistic action-at-a-distance models
have been proposed \cite{poincare} - \cite{vlad}, and compared with observations \cite{whitrow, will}.
A major difficulty with most of the models proposed is their disagreement with Einstein's theory of 
General Relativity (GR) in the so-called
"slow motion approximation" (first Post-Newtonian approximation (1PN)), even for the simpler case of two 
point masses ($N=2$) . 

The objective of this paper is to present a relativistic action-at-a-distance description of gravitational interactions
for a system consisting of an arbitrary number $N$ of point masses. 
  
The model presented in this paper is in agreement with GR at 1PN for an arbitrary number $N$ of interacting point masses.

Our description also agrees with GR in the so-called "fast motion approximation" for
$N$ point masses (at the first Post-Minkowskian order (1PM)) and it is in agreement with GR for
the one body case ($N=1$) at all orders (assuming the central mass is not spinning) if the Schwarzschild 
metric is expressed in the isotropic gauge.

{\it An action functional for gravity in the relativistic action-at-a-distance formulation}

In order to describe a relativistic system of $N$ point masses interacting
gravitationally, we consider the following action functional:

\begin{eqnarray}
S  & = & -\sum\limits^{}_{i} m_{i} c \int d\lambda_{i} \zeta_{i} 
+ \sum\limits^{}_{i} \sum\limits^{}_{j \neq i} \frac{G m_{i} m_{j}}{c} \int \int d\lambda_{i} d\lambda_{j}
\delta\left(\rho_{ij}\right)
F_{ij} \nonumber\\
& & + \sum\limits^{}_{i} \sum\limits^{}_{j \neq i} \sum\limits^{}_{k\neq i, j} \frac{G^{2} m_{i} m_{j} m_{k}}{c^{3}}
\int \int \int d\lambda_{i} d\lambda_{j} d\lambda_{k} \delta\left(\rho_{ij}\right) \delta\left(\rho_{j k}\right)
F_{ijk}  + ...
\label{1}
\end{eqnarray}

In (\ref{1}), $m_{i}$ $(i= 1,2,...,N)$ is the mass of particle $i$, 
$\lambda_{i}$ is a Poincar\'e invariant parameter labelling the events along
the world line $z^{\mu}_{i}(\lambda_{i})$ of particle $i$, $c$ is the speed of light and
$G$ the universal gravitational constant.

The functions
$F_{ij} = F_{ij} \left(\xi_{ij}, \gamma_{ij}, \gamma_{ji}, \zeta_{i}, \zeta_{j} \right)$, 
$F_{ijk} = F_{ijk} \left(\xi_{ij}, \xi_{ik}, \xi_{jk}, \gamma_{ij}, \gamma_{ji}, \gamma_{ik}, \gamma_{ki}, \gamma_{jk}, \gamma_{kj}, 
\zeta_{i}, \zeta_{j}, \zeta_{k} \right)$, are invariant under Poincar\'e transformations since they themselves are assumed to be 
functions of the Poincar\'e invariants $\xi_{i j}$, $\gamma_{i j}$, and $\zeta_{i}$.
The Poincar\'e invariants $\rho_{i j}$, $\xi_{i j}$, $\gamma_{i j}$, and $\zeta_{i}$, 
are defined as follows \cite{PLB}:

\begin{equation}
\rho_{i j}= \left( z_{i} - z_{j}\right)^{2} 
\label{2a}
\end{equation}

\begin{equation}
\xi_{i j}= \left(\dot{z}_{i} \dot{z}_{j}\right) 
\label{2b}
\end{equation}

\begin{equation}
\gamma_{i j}= \left(\dot{z}_{i} (z_{j} - z_{i})\right) 
\label{2c}
\end{equation}

\begin{equation}
\zeta_{i} = \dot{z}_{i}^{2}
\label{2d}
\end{equation}

We denote
$\dot{z}^{\mu}_{i} = \frac{d z^{\mu}_{i}}{d \lambda_{i}}$.
The metric tensor: $\eta_{\mu\nu}= diag(+1,-1,-1,-1)$.

The action functional (\ref{1})
is invariant under Lorentz transformations and does not involve any fields to mediate
the interactions between the masses. The particles interact with each other directly
and we assume that the interactions
propagate at the speed of light $c$ in vacuum. The Dirac delta functions in (\ref{1}) 
account for the interactions propagating at the speed
of light forward and backward in time.

The action (\ref{1}) can be written in a compact form as follows:

\begin{eqnarray}
S  & = & -\sum\limits^{}_{i} m_{i} c \int d\lambda_{i} \zeta_{i} \nonumber\\
& & + \sum\limits^{N}_{k=2} \sum\limits^{}_{i_{1}} 
\sum\limits^{}_{i_{2} \neq i_{1}} ... \sum\limits^{}_{i_{k} \neq i_{1}, ..., i_{k-1}}
\frac{G^{k-1} m_{i_{1}}... m_{i_{k}}}{c^{2k-3}} \int ...\int d\lambda_{i_{1}}... d\lambda_{i_{k}}
\prod^{k-1}_{l=1}
\delta\left(\rho_{i_{l}i_{l+1}}\right)
F_{i_{1}...i_{k}} 
\label{3}
\end{eqnarray}

Notice that not only the two-body interactions ($k=2$), but all possible k-body interactions ($k=2,...,N$) 
contribute to the action.

Without loss of generality we can assume that $F_{ji} = F_{ij}$, $F_{kji} = F_{ijk}$ and so on 
($F_{i_{k} i_{2}...i_{k-1} i_{1}} =F_{i_{1} i_{2}... i_{k-1} i_{k}}$).

From (\ref{1}), we can see that we can write the action of an individual particle $i$ as follows:

\begin{eqnarray}
& & S_{i}  =  - m_{i} c \int d\lambda_{i} 
\left(\zeta_{i} - \frac{2 G}{c^{2}} \sum\limits^{}_{j \neq i} m_{j} \int d\lambda_{j} \delta(\rho_{ij}) F_{ij}
\right. \nonumber\\
& & \left.  - \frac{G^{2}}{c^{4}} \sum\limits^{}_{j \neq i} \sum\limits^{}_{k \neq i, j} m_{j} m_{k}
\int \int d\lambda_{j} d\lambda_{k} \left(
\delta(\rho_{ij}) \delta(\rho_{jk}) F_{ijk} 
+ \delta(\rho_{jk}) \delta(\rho_{ki}) F_{jki}  + \delta(\rho_{ki}) \delta(\rho_{ij}) F_{kij} \right) 
 + ...\right) \nonumber\\
\label{6}
\end{eqnarray}

The equations of motion of the relativistic particles can be derived from the action (\ref{1}) (or from (\ref{6}))
using the variational principle. We find:

\begin{eqnarray}
& & \ddot{z}_{i}^{\mu} + \frac{G}{c^{2}} \sum\limits^{}_{j \neq i} m_{j} \int d\lambda_{j} \left( \frac{\partial}{\partial z_{i\mu}} \left(\delta(\rho_{ij}) F_{ij}\right)
- \frac{d}{d \lambda_{i}} \left(\delta(\rho_{ij}) \frac{\partial F_{ij}}{\partial \dot{z}_{i\mu}}\right)\right) \nonumber\\
& & + \frac{G^{2}}{2 c^{4}} \sum\limits^{}_{j \neq i} \sum\limits^{}_{k \neq i,j} m_{j} m_{k} \int \int d\lambda_{j} d\lambda_{k} 
\left(\delta(\rho_{jk}) 
\left( \frac{\partial}{\partial z_{i\mu}} \left(\delta(\rho_{ij}) F_{ijk}\right)
- \frac{d}{d \lambda_{i}} \left(\delta(\rho_{ij}) \frac{\partial F_{ijk}}{\partial \dot{z}_{i\mu}}\right)\right) \right.\nonumber\\
& & \left. + \delta(\rho_{jk}) 
\left( \frac{\partial}{\partial z_{i\mu}} \left(\delta(\rho_{ki}) F_{jki}\right)
- \frac{d}{d \lambda_{i}} \left(\delta(\rho_{ki}) \frac{\partial F_{jki}}{\partial \dot{z}_{i\mu}}\right)\right) \right. \nonumber\\
& & \left. + \frac{\partial}{\partial z_{i\mu}} \left(\delta(\rho_{ki}) \delta(\rho_{ij}) F_{kij}\right)
- \frac{d}{d \lambda_{i}} \left(\delta(\rho_{ki}) \delta(\rho_{ij}) \frac{\partial F_{kij}}{\partial \dot{z}_{i\mu}}\right)
\right) +...\nonumber\\
&  &  = 0.
\label{7}
\end{eqnarray}

Integrating by parts and taking into
account that:

\begin{equation}
\frac{d}{d \lambda_{i}} \left(\delta(\rho_{ij})\right) =
\frac{\left(\frac{d \rho_{ij}}{d \lambda_{i}}\right)}
{\left(\frac{d \rho_{ij}}{d \lambda_{j}}\right)} 
\frac{d}{d \lambda_{j}} \left(\delta(\rho_{ij})\right) =
\frac{\gamma_{ij}}{\gamma_{ji}}
\frac{d}{d \lambda_{j}} \left(\delta(\rho_{ij})\right),
\label{7a}
\end{equation}

\noindent we can write the equations of motion (\ref{7}) in the form:

\begin{eqnarray}
\ddot{z}^{\mu}_{i} & + &
\frac{G}{c^2} \sum\limits^{}_{j\neq i}  m_{j}
\int d\lambda_{j} \delta\left(\rho_{ij}\right)
\left(A^{\mu}_{i j} + B^{\mu \nu}_{i j} \ddot{z}_{i \nu} +
C^{\mu \nu}_{i j} \ddot{z}_{j \nu} \right)\nonumber\\
& + &
\frac{G^{2}}{c^{4}} \sum\limits^{}_{j \neq i} \sum\limits^{}_{k \neq i, j} m_{j} m_{k}
\int \int d\lambda_{j} d\lambda_{k} \delta(\rho_{ij}) \delta(\rho_{jk})
\left(A^{\mu}_{ijk} + B^{\mu \nu}_{ijk} \ddot{z}_{i \nu} +
C^{\mu \nu}_{ijk} \ddot{z}_{j \nu} + D^{\mu \nu}_{ijk} \ddot{z}_{k \nu}\right)\nonumber\\
& + & \frac{G^{2}}{2 c^{4}} \sum\limits^{}_{j \neq i} \sum\limits^{}_{k \neq i, j} m_{j} m_{k}
\int \int d\lambda_{j} d\lambda_{k} \delta(\rho_{ji}) \delta(\rho_{ik})
\left(\tilde{A}^{\mu}_{jik} + \tilde{B}^{\mu \nu}_{jik} \ddot{z}_{i \nu} +
\tilde{C}^{\mu \nu}_{jik} \ddot{z}_{j \nu} + \tilde{D}^{\mu \nu}_{jik} \ddot{z}_{k \nu}\right) + ...\nonumber\\
& &  = 0,
\label{5}
\end{eqnarray}

\noindent where

\begin{eqnarray}
A^{\mu}_{i j} & = & \frac{\partial F_{ij}}{\partial z_{i \mu}} -
\frac{\partial^{2} F_{ij}}{\partial z^{\eta}_{i}\partial \dot{z}_{i \mu}}
\dot{z}^{\eta}_{i} +
\frac{\zeta_{j}}{\gamma_{ji}^{2}} \left( (z^{\mu}_{i} - z^{\mu}_{j}) F_{ij} +
\gamma_{ij} \frac{\partial F_{ij}}{\partial\dot{z}_{i \mu}}\right) \nonumber\\
& & + \frac{1}{\gamma_{ji}} \left( - \dot{z}^{\mu}_{j} F_{ij} +
(z^{\mu}_{i} - z^{\mu}_{j}) \frac{\partial F_{ij}}{\partial z^{\eta}_{j}}
\dot{z}^{\eta}_{j} + \xi_{ij} \frac{\partial F_{ij}}{\partial \dot{z}_{i \mu}}
+ \gamma_{ij} \frac{\partial^{2} F_{ij}}{\partial z^{\eta}_{j}
\partial \dot{z}_{i \mu}} \dot{z}^{\eta}_{j}\right),
\label{6a}
\end{eqnarray}

\begin{equation}
B^{\mu \nu}_{i j} =
- \frac{\partial^{2} F_{ij}}{\partial \dot{z}_{i \mu} \partial \dot{z}_{i \nu}}, 
\label{6b}
\end{equation}

\begin{equation}
C^{\mu \nu}_{i j} =
\frac{(z^{\mu}_{i} -z^{\mu}_{j})}{\gamma_{ji}}
\left(\frac{\partial F_{ij}}{\partial \dot{z}_{j \nu}}
- \frac{(z^{\nu}_{i} -z^{\nu}_{j})}{\gamma_{ji}} F_{ij} \right)
+ \frac{\gamma_{ij}}{\gamma_{ji}} \left(
\frac{\partial^{2} F_{ij}}{\partial \dot{z}_{i \mu} \partial \dot{z}_{j \nu}}
- \frac{(z^{\nu}_{i} -z^{\nu}_{j})}{\gamma_{ji}}
\frac{\partial F_{ij}}{\partial \dot{z}_{i \mu}}
\right),
\label{6c}
\end{equation}

\begin{eqnarray}
A^{\mu}_{ijk} & = & \frac{\partial F_{ijk}}{\partial z_{i \mu}} -
\frac{\partial^{2} F_{ijk}}{\partial z^{\eta}_{i}\partial \dot{z}_{i \mu}}
\dot{z}^{\eta}_{i} +
\frac{\zeta_{j}}{\gamma_{ji}^{2}} \left( (z^{\mu}_{i} - z^{\mu}_{j}) F_{ijk} +
\gamma_{ij} \frac{\partial F_{ijk}}{\partial\dot{z}_{i \mu}}\right) \nonumber\\
& & + \frac{1}{\gamma_{ji}} \left( - \dot{z}^{\mu}_{j} F_{ijk} +
(z^{\mu}_{i} - z^{\mu}_{j}) \frac{\partial F_{ijk}}{\partial z^{\eta}_{j}}
\dot{z}^{\eta}_{j} + \xi_{ij} \frac{\partial F_{ijk}}{\partial \dot{z}_{i \mu}}
+ \gamma_{ij} \frac{\partial^{2} F_{ijk}}{\partial z^{\eta}_{j}
\partial \dot{z}_{i \mu}} \dot{z}^{\eta}_{j}\right) \nonumber\\
& & - \frac{1}{\gamma_{kj}\gamma_{ji}} \left(\xi_{jk} + \frac{\gamma_{jk}}{\gamma_{kj}} \zeta_{k}\right)
\left( (z^{\mu}_{i} - z^{\mu}_{j}) F_{ijk} +
\gamma_{ij} \frac{\partial F_{ijk}}{\partial\dot{z}_{i \mu}}\right) \nonumber\\
& & - \frac{\gamma_{jk}}{\gamma_{kj} \gamma_{ji}}
\left((z^{\mu}_{i} - z^{\mu}_{j}) \frac{\partial F_{ijk}}{\partial z^{\eta}_{k}}
\dot{z}^{\eta}_{k} + \gamma_{ij} \frac{\partial^{2} F_{ijk}}{\partial z^{\eta}_{k}
\partial \dot{z}_{i \mu}} \dot{z}^{\eta}_{k}\right),
\label{6aa}
\end{eqnarray}

\begin{equation}
B^{\mu \nu}_{ijk} =
- \frac{\partial^{2} F_{ijk}}{\partial \dot{z}_{i \mu} \partial \dot{z}_{i \nu}}, 
\label{6bb}
\end{equation}

\begin{equation}
C^{\mu \nu}_{ijk} =
\frac{(z^{\mu}_{i} -z^{\mu}_{j})}{\gamma_{ji}}
\left(\frac{\partial F_{ijk}}{\partial \dot{z}_{j \nu}}
- \frac{(z^{\nu}_{i} -z^{\nu}_{j})}{\gamma_{ji}} F_{ijk} \right)
+ \frac{\gamma_{ij}}{\gamma_{ji}} \left(
\frac{\partial^{2} F_{ijk}}{\partial \dot{z}_{i \mu} \partial \dot{z}_{j \nu}}
- \frac{(z^{\nu}_{i} -z^{\nu}_{j})}{\gamma_{ji}}
\frac{\partial F_{ijk}}{\partial \dot{z}_{i \mu}}
\right),
\label{6cc}
\end{equation}

\begin{equation}
D^{\mu \nu}_{ijk} =
\frac{\gamma_{jk}(z^{\nu}_{j} - z^{\nu}_{k})}{\gamma_{kj}^{2}} \left(
\frac{(z^{\mu}_{i} - z^{\mu}_{j})}{\gamma_{ji}}F_{ijk} + 
\frac{\gamma_{ij}}{\gamma_{ji}}\frac{\partial F_{ijk}}{\partial \dot{z}_{i \mu}}\right)
- \frac{\gamma_{jk}}{\gamma_{kj}} \left(
\frac{(z^{\mu}_{i} - z^{\mu}_{j})}{\gamma_{ji}}\frac{\partial F_{ijk}}{\partial \dot{z}_{k \nu}} + 
\frac{\gamma_{ij}}{\gamma_{ji}}\frac{\partial^{2} F_{ijk}}{\partial \dot{z}_{i \mu} \partial \dot{z}_{k \nu}}\right),
\label{6dd}
\end{equation}

\begin{eqnarray}
\tilde{A}^{\mu}_{jik} & = & \frac{\partial F_{jik}}{\partial z_{i \mu}} -
\frac{\partial^{2} F_{jik}}{\partial z^{\eta}_{i}\partial \dot{z}_{i \mu}}
\dot{z}^{\eta}_{i} +
\frac{\zeta_{j}}{\gamma_{ji}^{2}} \left( (z^{\mu}_{i} - z^{\mu}_{j}) F_{jik} +
\gamma_{ij} \frac{\partial F_{jik}}{\partial\dot{z}_{i \mu}}\right) 
+ \frac{\zeta_{k}}{\gamma_{ki}^{2}} \left( (z^{\mu}_{i} - z^{\mu}_{k}) F_{jik} +
\gamma_{ik} \frac{\partial F_{jik}}{\partial\dot{z}_{i \mu}}\right)\nonumber\\
& & + \frac{1}{\gamma_{ji}} \left( - \dot{z}^{\mu}_{j} F_{jik} +
(z^{\mu}_{i} - z^{\mu}_{j}) \frac{\partial F_{jik}}{\partial z^{\eta}_{j}}
\dot{z}^{\eta}_{j} + \xi_{ij} \frac{\partial F_{jik}}{\partial \dot{z}_{i \mu}}
+ \gamma_{ij} \frac{\partial^{2} F_{jik}}{\partial z^{\eta}_{j}
\partial \dot{z}_{i \mu}} \dot{z}^{\eta}_{j}\right) \nonumber\\
& & + \frac{1}{\gamma_{ki}} \left( - \dot{z}^{\mu}_{k} F_{jik} +
(z^{\mu}_{i} - z^{\mu}_{k}) \frac{\partial F_{jik}}{\partial z^{\eta}_{k}}
\dot{z}^{\eta}_{k} + \xi_{ik} \frac{\partial F_{jik}}{\partial \dot{z}_{i \mu}}
+ \gamma_{ik} \frac{\partial^{2} F_{jik}}{\partial z^{\eta}_{k}
\partial \dot{z}_{i \mu}} \dot{z}^{\eta}_{k}\right),
\label{6aaa}
\end{eqnarray}

\begin{equation}
\tilde{B}^{\mu \nu}_{jik} =
- \frac{\partial^{2} F_{jik}}{\partial \dot{z}_{i \mu} \partial \dot{z}_{i \nu}}, 
\label{6bbb}
\end{equation}

\begin{equation}
\tilde{C}^{\mu \nu}_{jik} =
\frac{(z^{\mu}_{i} -z^{\mu}_{j})}{\gamma_{ji}}
\left(\frac{\partial F_{jik}}{\partial \dot{z}_{j \nu}}
- \frac{(z^{\nu}_{i} -z^{\nu}_{j})}{\gamma_{ji}} F_{jik} \right)
+ \frac{\gamma_{ij}}{\gamma_{ji}} \left(
\frac{\partial^{2} F_{jik}}{\partial \dot{z}_{i \mu} \partial \dot{z}_{j \nu}}
- \frac{(z^{\nu}_{i} -z^{\nu}_{j})}{\gamma_{ji}}
\frac{\partial F_{jik}}{\partial \dot{z}_{i \mu}}
\right),
\label{6ccc}
\end{equation}

\begin{equation}
\tilde{D}^{\mu \nu}_{jik} =
- \frac{(z^{\nu}_{i} - z^{\nu}_{k})}{\gamma_{ki}^{2}} \left(
(z^{\mu}_{i} - z^{\mu}_{k})F_{jik} + 
\gamma_{ik}\frac{\partial F_{jik}}{\partial \dot{z}_{i \mu}}\right)
+ \frac{1}{\gamma_{ki}} \left(
(z^{\mu}_{i} - z^{\mu}_{k})\frac{\partial F_{jik}}{\partial \dot{z}_{k \nu}} + 
\gamma_{ik}\frac{\partial^{2} F_{jik}}{\partial \dot{z}_{i \mu} \partial \dot{z}_{k \nu}}\right).
\label{6ddd}
\end{equation}

Multiplying (\ref{7}) by $\dot{z}_{i\mu}$ (and performing the summation over $\mu$), we find that the solutions of the equations of motion (\ref{7})
must satisfy the following $N$ conditions ($i=1,2,...,N$):

\begin{eqnarray}
& & \frac{d}{d \lambda_{i}} \left(
\zeta_{i} + \frac{2 G}{c^{2}} \sum\limits^{}_{j \neq i} m_{j} \int d\lambda_{j} \delta(\rho_{ij}) \left(F_{ij} - \dot{z}_{i}^{\mu} 
\frac{\partial F_{ij}}{\partial \dot{z}_{i}^{\mu}}\right) 
\right. \nonumber\\
& &  + \frac{G^{2}}{c^{4}} \sum\limits^{}_{j \neq i} \sum\limits^{}_{k \neq i, j} m_{j} m_{k}
\int \int d\lambda_{j} d\lambda_{k} \left(
\delta(\rho_{ij}) \delta(\rho_{jk}) 
\left(F_{ijk} - \dot{z}_{i}^{\mu} 
\frac{\partial F_{ijk}}{\partial \dot{z}_{i}^{\mu}}\right) \right. \nonumber\\
& & \left. \left. + \delta(\rho_{jk}) \delta(\rho_{ki}) \left(F_{jki} - \dot{z}_{i}^{\mu} 
\frac{\partial F_{jki}}{\partial \dot{z}_{i}^{\mu}}\right) 
+ \delta(\rho_{ki}) \delta(\rho_{ij}) \left(F_{kij} - \dot{z}_{i}^{\mu} 
\frac{\partial F_{kij}}{\partial \dot{z}_{i}^{\mu}}\right) \right)
 + ...\right) \nonumber\\
&  & = 0.
\label{8}
\end{eqnarray}

Let us assume that $F_{i_{1}... i_{k}}$ ($k = 2,...,N$) are homogeneous functions of degree two in
$\dot{z}_{i_{1}}$, ..., $\dot{z}_{i_{k}}$, i.e., we assume that they satisfy the following
conditions:

\begin{equation}
\dot{z}^{\mu}_{i_{1}} \frac{\partial F_{i_{1} ... i_{k}}}{\partial \dot{z}^{\mu}_{i_{1}}}
= ... = \dot{z}^{\mu}_{i_{k}} \frac{\partial F_{i_{1} ... i_{k}}}{\partial \dot{z}^{\mu}_{i_{k}}}
= 2 F_{i_{1} ... i_{k}}
\label{4}
\end{equation}

The conditions (\ref{8}) combined with (\ref{4}) guarantee that, for the solutions of the equations of motion, the expressions:

\begin{eqnarray}
\zeta_{i} & - & \frac{2 G}{c^{2}} \sum\limits^{}_{j \neq i} m_{j} \int d\lambda_{j} \delta(\rho_{ij}) F_{ij}\nonumber\\
& - & \frac{G^{2}}{c^{4}} \sum\limits^{}_{j \neq i} \sum\limits^{}_{k \neq i, j} m_{j} m_{k}
\int \int d\lambda_{j} d\lambda_{k} \left(
\delta(\rho_{ij}) \delta(\rho_{jk}) F_{ijk} + \delta(\rho_{jk}) \delta(\rho_{ki}) F_{jki}  
+ \delta(\rho_{ki}) \delta(\rho_{ij}) F_{kij} \right) + ...\nonumber\\ 
& = & c_{i},
\label{8aa}
\end{eqnarray}

\noindent are constants (which by simple scaling can be made equal to $1$):

\begin{eqnarray}
\zeta_{i} & - & \frac{2 G}{c^{2}} \sum\limits^{}_{j \neq i} m_{j} \int d\lambda_{j} \delta(\rho_{ij}) F_{ij}\nonumber\\
& - & \frac{G^{2}}{c^{4}} \sum\limits^{}_{j \neq i} \sum\limits^{}_{k \neq i, j} m_{j} m_{k}
\int \int d\lambda_{j} d\lambda_{k} \left(
\delta(\rho_{ij}) \delta(\rho_{jk}) F_{ijk} + \delta(\rho_{jk}) \delta(\rho_{ki}) F_{jki}  
+ \delta(\rho_{ki}) \delta(\rho_{ij}) F_{kij} \right) + ...\nonumber\\ 
& = & 1.
\label{8a}
\end{eqnarray}

From (\ref{4}) it immediately follows that:

\begin{equation}
F_{i_{1} ... i_{k}} = \frac{1}{2} \frac{\partial^{2} F_{i_{1} ... i_{k}}}{\partial \dot{z}^{\mu}_{i_{1}} \partial \dot{z}^{\nu}_{i_{1}}} 
\dot{z}^{\mu}_{i_{1}} \dot{z}^{\nu}_{i_{1}}
= ... = \frac{1}{2} \frac{\partial^{2} F_{i_{1} ... i_{k}}}{\partial \dot{z}^{\mu}_{i_{k}} \partial \dot{z}^{\nu}_{i_{k}}} 
\dot{z}^{\mu}_{i_{k}} \dot{z}^{\nu}_{i_{k}}
\label{5a}
\end{equation}

Using (\ref{5a}) we immediately see that the action for particle $i$ (\ref{6}) can be rewritten in a compact form as:

\begin{equation}
S_{i} = - m_{i} c \int d\lambda_{i} g^{(i)}_{\mu\nu} \dot{z}_{i}^{\mu} \dot{z}_{i}^{\nu}
\label{9}
\end{equation}

\noindent where,

\begin{eqnarray}
g^{(i)}_{\mu\nu} & = & \eta_{\mu\nu} - \frac{G}{c^{2}} \sum\limits^{}_{j \neq i} m_{j} \int d\lambda_{j} \delta(\rho_{ij}) 
\frac{\partial^{2}F_{ij}}{\partial \dot{z}_{i}^{\mu} \partial \dot{z}_{i}^{\nu}}
\nonumber\\
& &  - \frac{G^{2}}{2 c^{4}} \sum\limits^{}_{j \neq i} \sum\limits^{}_{k \neq i, j} m_{j} m_{k}
\int \int d\lambda_{j} d\lambda_{k} \left(
\delta(\rho_{ij}) \delta(\rho_{jk}) 
\frac{\partial^{2}F_{ijk}}{\partial \dot{z}_{i}^{\mu} \partial \dot{z}_{i}^{\nu}} 
+ \delta(\rho_{jk}) \delta(\rho_{ki}) 
\frac{\partial^{2}F_{jki}}{\partial \dot{z}_{i}^{\mu} \partial \dot{z}_{i}^{\nu}}  \right. \nonumber\\
& & \left. + \delta(\rho_{ki}) \delta(\rho_{ij}) \frac{\partial^{2}F_{kij}}{\partial \dot{z}_{i}^{\mu} \partial \dot{z}_{i}^{\nu}} \right) 
+ ...
\label{10}
\end{eqnarray}

From (\ref{10}, \ref{4}, \ref{5a}) it follows that:

\begin{equation}
\frac{\partial g^{(i)}_{\alpha\beta}}{\partial \dot{z}_{i}^{\mu}} \dot{z}_{i}^{\alpha} = 0,
\label{11a}
\end{equation}

\begin{equation}
\frac{\partial^{2} g^{(i)}_{\alpha\beta}}{\partial \dot{z}_{i}^{\mu} \partial \dot{z}_{i}^{\nu}} 
\dot{z}_{i}^{\alpha} \dot{z}_{i}^{\beta} = 0,
\label{11aa}
\end{equation}

\begin{equation}
\frac{\partial^{2} g^{(i)}_{\alpha\beta}}{\partial \dot{z}_{i}^{\mu} \partial z_{i}^{\nu}} 
\dot{z}_{i}^{\alpha} = 0.
\label{11aaa}
\end{equation}

Using (\ref{10}, \ref{11a}, \ref{11aa}, \ref{11aaa}) the equations of motion (\ref{7}) can also be written 
in a more compact form:

\begin{equation}
g^{(i)}_{\mu\nu}  \ddot{z}_{i}^{\nu} + \frac{1}{2} \left(
\frac{\partial g^{(i)}_{\mu\alpha}}{\partial z_{i}^{\beta}} + \frac{\partial g^{(i)}_{\mu\beta}}{\partial z_{i}^{\alpha}}
- \frac{\partial g^{(i)}_{\alpha\beta}}{\partial z_{i}^{\mu}}
\right) \dot{z}_{i}^{\alpha}\dot{z}_{i}^{\beta} = 0
\label{11}
\end{equation}

From (\ref{5}, \ref{8}, \ref{10}), we find that the conditions (\ref{8}) can be simply expressed as:

\begin{equation}
\frac{d}{d\lambda_{i}} \left(g^{(i)}_{\mu\nu} \dot{z}_{i}^{\mu}\dot{z}_{i}^{\nu}\right) = 0
\label{12}
\end{equation}

From (\ref{8a}, \ref{5a},\ref{10}) it follows that for the solutions of the equations of motion:

\begin{equation}
d\lambda_{i}^{2} = g^{(i)}_{\mu\nu} dz_{i}^{\mu} dz_{i}^{\nu}
\label{12a}
\end{equation}

Notice that $g^{(i)}_{\mu\nu}$ is not a field. It depends not only on $z_{i}$, $z_{j}$ and $\dot{z}_{j}$ ($j \ne i$), 
but also on $\dot{z}_{i}$!

The main task in our formulation is to determine the functions $F_{ij}$, $F_{ijk}$, etc, in (\ref{3}), and to verify that the
predictions of the theory are in agreement with observations.

{\it Test particles and the formulation of the action-at-a-distance model as a field theory}

Let us assume that in the limit $m_{i} \to 0$ the tensor $g^{(i)}_{\mu\nu}$ does not depend on $\dot{z}_{i}$. Only in this limit,
in which $m_{i}$ is a test particle, we may have a field interpretation for the metric tensor $g^{(i)}_{\mu\nu}$.

Let us consider a system of $N+1$ point particles, one of them being a test particle of mass $m$ and the other $N$ particles
having masses $m_{i}$ ($i = 1,...,N$). Let $z(\lambda)$ be the worldline of the test particle. From (\ref{9}) we see that 
we can write the action for the test particle as follows:

\begin{equation}
S = - m c \int d\lambda g_{\mu\nu} \dot{z}^{\mu} \dot{z}^{\nu}.
\label{13}
\end{equation}

In (\ref{13}), the metric tensor $g_{\mu\nu}$ depends on $z$ (and on $z_{i}$ and $\dot{z}_{i}$ ($i=1,...,N$)), but does 
not depend on $\dot{z}$. It can be given a field interpretation, if one desires to do so\cite{domingo3}.

From (\ref{11}) we see that for a test particle the equations of motion are:

\begin{equation}
g_{\mu\nu} \ddot{z}^{\nu} + \frac{1}{2} \left(
\frac{\partial g_{\mu\alpha}}{\partial z^{\beta}} + \frac{\partial g_{\mu\beta}}{\partial z^{\alpha}}
- \frac{\partial g_{\alpha\beta}}{\partial z^{\mu}}
\right) \dot{z}^{\alpha}\dot{z}^{\beta} = 0
\label{14}
\end{equation}

Assuming that the matrix $g_{\mu\nu}$ is invertible, these are, of course, the well known equations for geodesics.

It may be possible to impose conditions on the functions $F_{ij}$, $F_{ijk}$, etc, in (\ref{3}) if, for example,
one demands that the metric tensor $g_{\mu\nu}$ (which is associated with a test particle) obeys Einstein's field equations.
Of course, there is no guarantee that this can be done at all orders in the Post-Minkowskian expansion, 
either due to the mathematical complexity of the equations or to
the possibility that Einstein's theory of General Relativity may not exactly admit a dual (action-at-a-distance) formulation,
or at least not one described by an action of the form (\ref{3}). At the first Post-Minkowskian order it is known that
$F_{ij} = 2\xi_{ij}^{2} - \zeta_{ij}^{2}$ \cite{golberg} - \cite{friedman}.

{\it A possible expression for $F_{ij}$}

Let us assume that the functions $F_{ij}$ can be expressed as follows:

\begin{equation}
F_{ij} = \alpha\left(\epsilon_{ij},\epsilon_{ji}\right) \xi_{ij}^{2} + \beta\left(\epsilon_{ij},\epsilon_{ji}\right) \zeta_{ij}^{2},
\label{15}
\end{equation}

\noindent where,

\begin{equation}
\epsilon_{ij} = \frac{G m_{i}}{2 c^{2} |\eta_{ij}|},
\label{16}
\end{equation}

and\cite{PLB}:

\begin{equation}
\zeta_{ij} = \sqrt{\zeta_{i}\zeta_{j}},
\label{17}
\end{equation}

\begin{equation}
\eta_{ij} = \frac{\gamma_{ij}}{\sqrt{\zeta_{i}}}.
\label{18}
\end{equation}

We assume that the functions $\alpha$ and $\beta$ are symmetric:

\begin{equation}
\alpha\left(\epsilon_{ij},\epsilon_{ji}\right) = \alpha\left(\epsilon_{ji},\epsilon_{ij}\right),
\label{19a}
\end{equation}

\begin{equation}
\beta\left(\epsilon_{ij},\epsilon_{ji}\right) = \beta\left(\epsilon_{ji},\epsilon_{ij}\right).
\label{19b}
\end{equation}

{\it The one-body problem}

Let us consider the case of a test particle interacting with a particle of mass $M$. This is the case $N = 1$ (the one-body problem).
The motion of the mass $M$ is not affected by the presence of the test particle. The mass $M$ moves with constant velocity in any
inertial reference frame.

Let us, for simplicity, consider the inertial frame in which the mass $M$ is at rest and positioned at the origin of the coordinate system.
In this frame of reference the world line of the test particle is described by the four-vector $z^{\mu} = (ct, \vec{r})$.
From (\ref{10}) and (\ref{15}) we find the components of the metric tensor $g_{\mu\nu}$ in this reference frame to be as follows:

\begin{equation}
g_{00} = 1 - \frac{2GM \left(\alpha(0,\epsilon) + \beta(0,\epsilon)\right)}{c^{2} r},
\label{20a}
\end{equation}

\begin{equation}
g_{0i} = 0,
\label{20b}
\end{equation}

\begin{equation}
g_{ij} = - \delta_{ij} \left(1 - \frac{2GM \beta(0,\epsilon)}{c^{2} r}\right).
\label{20c}
\end{equation}

In (\ref{20a} - \ref{20c}):

\begin{equation}
\epsilon = \frac{GM}{2 c^{2} r}.
\label{21}
\end{equation}

If we choose the functions $\alpha$ and $\beta$ as follows:

\begin{equation}
\alpha(0,\epsilon) = \frac{(1 + \epsilon)^{4} - \frac{(1 - \epsilon)^{2}}{(1 + \epsilon)^{2}}}{4\epsilon},
\label{22a}
\end{equation}

\begin{equation}
\beta(0,\epsilon) = \frac{1 - (1 + \epsilon)^{4}}{4\epsilon},
\label{22b}
\end{equation}

\noindent one can easily check that the metric (\ref{20a} - \ref{20c}), with $\alpha$ and $\beta$ given by (\ref{22a},\ref{22b}), coincides
with the well known Schwarzschild metric of GR in isotropic form\cite{weinberg}.

Since $\alpha$, $\beta$ and $\epsilon_{ji}$ are Poincar\'e invariants, we can write the functional relations:

\begin{equation}
\alpha(0,\epsilon_{ji}) = \frac{(1 + \epsilon_{ji})^{4} - \frac{(1 - \epsilon_{ji})^{2}}{(1 + \epsilon_{ji})^{2}}}{4\epsilon_{ji}},
\label{23a}
\end{equation}

\begin{equation}
\beta(0,\epsilon_{ji}) = \frac{1 - (1 + \epsilon_{ji})^{4}}{4\epsilon_{ji}}.
\label{23b}
\end{equation}

At the second Post-Minkowskian order (up to terms proportional to $G^{2}$ in the metric), we can write:

\begin{equation}
\alpha(0,\epsilon_{ji}) \approx \alpha_{0} + \alpha_{1} \epsilon_{ji},
\label{24a}
\end{equation}

\begin{equation}
\beta(0,\epsilon_{ji}) \approx \beta_{0} + \beta_{1} \epsilon_{ji},
\label{24b}
\end{equation}

\noindent where $\alpha_{0}$, $\beta_{0}$, $\alpha_{1}$ and $\beta_{1}$ are constants.

The values of these constants can easily be determined by expanding (\ref{23a}) and (\ref{23b}). We find:

\begin{equation}
\alpha_{0} = 2,
\label{25a}
\end{equation}

\begin{equation}
\beta_{0} = -1,
\label{25b}
\end{equation}

\begin{equation}
\alpha_{1} = - \frac{1}{2},
\label{25c}
\end{equation}

\begin{equation}
\beta_{1} = - \frac{3}{2}.
\label{25d}
\end{equation}

{\it The second Post-Minkowskian approximation}

Let us now consider the gravitational $N$-body problem described by the action (\ref{3}). 
Assume that $F_{ij}$ are given by (\ref{15}, \ref{19a},\ref{19b}). At the second Post-Minkowskian (2PM)
order the functions $\alpha$ and $\beta$ will be given by the expressions:

\begin{equation}
\alpha(\epsilon_{ij},\epsilon_{ji}) \approx \alpha_{0} + \alpha_{1} \left(\epsilon_{ij} + \epsilon_{ji}\right),
\label{26a}
\end{equation}

\begin{equation}
\beta(\epsilon_{ij},\epsilon_{ji}) \approx \beta_{0} + \beta_{1} \left(\epsilon_{ij} + \epsilon_{ji}\right).
\label{26b}
\end{equation}

Let us consider the case where the functions $F_{ijk}$ can be written as:

\begin{eqnarray}
F_{ijk} & = & a(\epsilon_{ij},\epsilon_{ji},\epsilon_{ki},\epsilon_{ik},\epsilon_{jk},\epsilon_{kj}) \xi_{ij}\xi_{jk}\xi_{ki}\nonumber\\
& & + b(\epsilon_{ij},\epsilon_{ji},\epsilon_{ki},\epsilon_{ik},\epsilon_{jk},\epsilon_{kj}) \zeta_{i}\zeta_{j}\zeta_{k}\nonumber\\
& & + c(\epsilon_{ij},\epsilon_{ji},\epsilon_{ki},\epsilon_{ik},\epsilon_{jk},\epsilon_{kj}) \xi_{ki}^{2}\zeta_{j}.
\label{27}
\end{eqnarray}

We assume that the functions $a$, $b$ and $c$ are symmetric in the indexes $(ik)$:

\begin{equation}
a(\epsilon_{kj},\epsilon_{jk},\epsilon_{ik},\epsilon_{ki},\epsilon_{ji},\epsilon_{ij})
= a(\epsilon_{ij},\epsilon_{ji},\epsilon_{ki},\epsilon_{ik},\epsilon_{jk},\epsilon_{kj}),
\label{28a} 
\end{equation}

\begin{equation}
b(\epsilon_{kj},\epsilon_{jk},\epsilon_{ik},\epsilon_{ki},\epsilon_{ji},\epsilon_{ij})
= b(\epsilon_{ij},\epsilon_{ji},\epsilon_{ki},\epsilon_{ik},\epsilon_{jk},\epsilon_{kj}),
\label{28b}
\end{equation}

\begin{equation}
c(\epsilon_{kj},\epsilon_{jk},\epsilon_{ik},\epsilon_{ki},\epsilon_{ji},\epsilon_{ij})
= c(\epsilon_{ij},\epsilon_{ji},\epsilon_{ki},\epsilon_{ik},\epsilon_{jk},\epsilon_{kj}).
\label{28c}
\end{equation}

At 2PM we have:

\begin{equation}
a(\epsilon_{ij},\epsilon_{ji},\epsilon_{ki},\epsilon_{ik},\epsilon_{jk},\epsilon_{kj}) \approx a_{0},
\label{29a} 
\end{equation}

\begin{equation}
b(\epsilon_{ij},\epsilon_{ji},\epsilon_{ki},\epsilon_{ik},\epsilon_{jk},\epsilon_{kj}) \approx b_{0},
\label{29b} 
\end{equation}

\begin{equation}
c(\epsilon_{ij},\epsilon_{ji},\epsilon_{ki},\epsilon_{ik},\epsilon_{jk},\epsilon_{kj}) \approx c_{0},
\label{29c} 
\end{equation}

\noindent where $a_{0}$, $b_{0}$ and $c_{0}$ are constants.

Therefore, at the second Post-Minkowskian order we can write the action, for a system of $N$ particles interacting gravitationally,
as follows:

\begin{eqnarray}
S  & = & -\sum\limits^{}_{i} m_{i} c \int d\lambda_{i} \zeta_{i} 
+ \sum\limits^{}_{i} \sum\limits^{}_{j \neq i} \frac{G m_{i} m_{j}}{c} \int \int d\lambda_{i} d\lambda_{j}
\delta\left(\rho_{ij}\right)
F_{ij} \nonumber\\
& & + \sum\limits^{}_{i} \sum\limits^{}_{j \neq i} \sum\limits^{}_{k\neq i, j} \frac{G^{2} m_{i} m_{j} m_{k}}{c^{3}}
\int \int \int d\lambda_{i} d\lambda_{j} d\lambda_{k} \delta\left(\rho_{ij}\right) \delta\left(\rho_{j k}\right)
F_{ijk}, 
\label{30}
\end{eqnarray}

\noindent where,

\begin{equation}
F_{ij} = \left(\alpha_{0} + \alpha_{1} \left(\epsilon_{ij} + \epsilon_{ji}\right)\right) \xi_{ij}^{2}
+ \left(\beta_{0} + \beta_{1} \left(\epsilon_{ij} + \epsilon_{ji}\right)\right) \zeta_{ij}^{2}
\label{31}
\end{equation}

\begin{equation}
F_{ijk} = a_{0} \xi_{ij}\xi_{jk}\xi_{ki}
+ b_{0} \zeta_{i}\zeta_{j}\zeta_{k}
+ c_{0} \xi_{ki}^{2}\zeta_{j}
\label{32}
\end{equation}

At the second Post-Minkowskian approximation there is no need to consider the functions $F_{i_{1}...i_{k}}$ for $k > 3$
since the terms associated with these functions in the action (\ref{3}) give contributions only at the ($k-1$)-Post-Minkowskian order.

At the second Post-Minkowskian order (2PM) the equations of motion are:

\begin{eqnarray}
\ddot{z}^{\mu}_{i} & + &
\frac{G}{c^2} \sum\limits^{}_{j\neq i}  m_{j}
\int d\lambda_{j} \delta\left(\rho_{ij}\right)
\left(A^{(0)\mu}_{i j} + A^{(1)\mu}_{i j} + B^{(0)\mu \nu}_{i j} \ddot{z}_{i \nu} +
C^{(0)\mu \nu}_{i j} \ddot{z}_{j \nu} \right)\nonumber\\
& + &
\frac{G^{2}}{c^{4}} \sum\limits^{}_{j \neq i} \sum\limits^{}_{k \neq i, j} m_{j} m_{k}
\int \int d\lambda_{j} d\lambda_{k} \delta(\rho_{ij}) \delta(\rho_{jk})
A^{(0)\mu}_{ijk} \nonumber\\
& + & \frac{G^{2}}{2 c^{4}} \sum\limits^{}_{j \neq i} \sum\limits^{}_{k \neq i, j} m_{j} m_{k}
\int \int d\lambda_{j} d\lambda_{k} \delta(\rho_{ji}) \delta(\rho_{ik})
\tilde{A}^{(0)\mu}_{jik} \nonumber\\
& = & 0.
\label{33}
\end{eqnarray}

Substituting (\ref{31}, \ref{32}) into (\ref{6a} - \ref{6c}) and (\ref{6aa}, \ref{6aaa}) we find (in this approximation):

\begin{eqnarray}
A^{(0)\mu}_{i j} & = & \frac{\left(z_{i}^{\mu} - z_{j}^{\mu}\right)}{\eta_{ji}^{2}}\left(
\alpha_{0} \xi_{ij}^{2} + \beta_{0} \zeta_{ij}^{2}\right)
+ \frac{2\dot{z}_{i}^{\mu}}{\zeta_{i}^{\frac{1}{2}} \eta_{ji}} 
\zeta_{ij}^{2} \left(\frac{\xi_{ij}}{\zeta_{ij}} + \frac{\eta_{ij}}{\eta_{ji}}\right) \beta_{0}\nonumber\\
& + & \frac{\dot{z}_{j}^{\mu}}{\zeta_{j}^{\frac{1}{2}} \eta_{ji}} \left[
\alpha_{0}\left(\xi_{ij}^{2} + 2\xi_{ij}\zeta_{ij}\frac{\eta_{ij}}{\eta_{ji}}\right) - \beta_{0}\zeta_{ij}^{2}\right],
\label{34a}
\end{eqnarray}

\begin{eqnarray}
A^{(1)\mu}_{i j} & = & \frac{2\left(z_{i}^{\mu} - z_{j}^{\mu}\right)}{\eta_{ji}^{2}}
\left(\epsilon_{ji} - \epsilon_{ij} \left(1 + \frac{\xi_{ij}\eta_{ij}}{\zeta_{ij}\eta_{ji}} - \frac{\eta_{ij}^{2}}{\eta_{ji}^{2}}\right)\right)
\left(\alpha_{1} \xi_{ij}^{2} + \beta_{1} \zeta_{ij}^{2}\right)\nonumber\\
& + & \frac{\dot{z}_{i}^{\mu}}{\zeta_{i}^{\frac{1}{2}} \eta_{ji}} \left[
4 \epsilon_{ji} \zeta_{ij}^{2} \left(\frac{\xi_{ij}}{\zeta_{ij}} + \frac{\eta_{ij}}{\eta_{ji}}\right) \beta_{1}
- \epsilon_{ij} \left(1 - \frac{\eta_{ij}^{2}}{\eta_{ji}^{2}}\right)  \left(\alpha_{1} \xi_{ij}^{2} + 3\beta_{1} \zeta_{ij}^{2}\right)\right]\nonumber\\
& + & \frac{2\dot{z}_{j}^{\mu}}{\zeta_{j}^{\frac{1}{2}} \eta_{ji}} \left[
\epsilon_{ji} \left(\alpha_{1}\left(2\xi_{ij}\zeta_{ij}\frac{\eta_{ij}}{\eta_{ji}} + \xi_{ij}^{2}\right) - \beta_{1}\zeta_{ij}^{2}\right)\right.\nonumber\\
& & \left. - \epsilon_{ij}\left(\alpha_{1}\left(\xi_{ij}^{2} + \xi_{ij}\zeta_{ij}\left(\frac{\eta_{ji}}{\eta_{ij}} - \frac{\eta_{ij}}{\eta_{ji}}\right)\right)
+ \beta_{1}\zeta_{ij}^{2}\right)\right],
\label{34aa}
\end{eqnarray}

\begin{equation}
B^{(0)\mu\nu}_{i j} = - 2\alpha_{0} \dot{z}_{j}^{\mu}\dot{z}_{j}^{\nu} - 2\beta_{0} \zeta_{j} \eta^{\mu\nu},
\label{34b}
\end{equation}

\begin{eqnarray}
C^{(0)\mu\nu}_{i j} & = & - \frac{\left(z_{i}^{\mu} - z_{j}^{\mu}\right)\left(z_{i}^{\nu} - z_{j}^{\nu}\right)}{\zeta_{j}\eta_{ji}^{2}}
\left(\alpha_{0} \xi_{ij}^{2} + \beta_{0} \zeta_{ij}^{2}\right)
+ \frac{2\left(z_{i}^{\mu} - z_{j}^{\mu}\right)}{\zeta_{j}^{\frac{1}{2}}\eta_{ji}}
\left(\alpha_{0}\xi_{ij}\dot{z}_{i}^{\nu} + \beta_{0}\zeta_{i}\dot{z}_{j}^{\nu}\right)\nonumber\\
& & - \frac{2\left(z_{i}^{\nu} - z_{j}^{\nu}\right)\zeta_{i}^{\frac{1}{2}}\eta_{ij}}{\zeta_{j}\eta_{ji}^{2}}
\left(\alpha_{0}\xi_{ij}\dot{z}_{j}^{\mu} + \beta_{0}\zeta_{j}\dot{z}_{i}^{\mu}\right)\nonumber\\
& & + 2\frac{\eta_{ij}\zeta_{i}^{\frac{1}{2}}}{\eta_{ji}\zeta_{j}^{\frac{1}{2}}}
\left(\alpha_{0}\left(\eta^{\mu\nu}\xi_{ij} + \dot{z}_{j}^{\mu}\dot{z}_{i}^{\nu}\right) + 2\beta_{0}\dot{z}_{i}^{\mu}\dot{z}_{j}^{\nu}\right),
\label{34c}
\end{eqnarray}

\begin{eqnarray}
A^{(0)\mu}_{ijk} & = & \frac{\left(z_{i}^{\mu} - z_{j}^{\mu}\right)}{\gamma_{ji}}
\left(\frac{\zeta_{j}}{\gamma_{ji}} - \frac{1}{\gamma_{kj}}\left(\xi_{jk} + \frac{\gamma_{jk}}{\gamma_{kj}}\zeta_{k}\right)\right)
\left(a_{0} \xi_{ij}\xi_{jk}\xi_{ki} + b_{0} \zeta_{i}\zeta_{j}\zeta_{k} + c_{0} \xi_{ki}^{2}\zeta_{j}\right)\nonumber\\
& & + \frac{2 \dot{z}_{i}^{\mu}}{\gamma_{ji}}\left(\zeta_{j}\frac{\gamma_{ij}}{\gamma_{ji}} + \xi_{ij} - \frac{\gamma_{ij}}{\gamma_{kj}}
\left(\xi_{jk} + \frac{\gamma_{jk}}{\gamma_{kj}}\zeta_{k}\right)
\right)\zeta_{j}\zeta_{k} b_{0}\nonumber\\
& & +\frac{\dot{z}_{j}^{\mu}}{\gamma_{ji}}\left(\left(\zeta_{j}\frac{\gamma_{ij}}{\gamma_{ji}} - \frac{\gamma_{ij}}{\gamma_{kj}}
\left(\xi_{jk} + \frac{\gamma_{jk}}{\gamma_{kj}}\zeta_{k}\right)\right)
\xi_{jk}\xi_{ki} a_{0} - \zeta_{i}\zeta_{j}\zeta_{k} b_{0} - \xi_{ki}^{2}\zeta_{j} c_{0}\right)\nonumber\\
& & + \frac{\dot{z}_{k}^{\mu}}{\gamma_{ji}}\left(\zeta_{j}\frac{\gamma_{ij}}{\gamma_{ji}} + \xi_{ij} - \frac{\gamma_{ij}}{\gamma_{kj}}
\left(\xi_{jk} + \frac{\gamma_{jk}}{\gamma_{kj}}\zeta_{k}\right)\right)\left(a_{0}\xi_{jk}\xi_{ij} + 2 c_{0}\xi_{ki}\zeta_{j}\right),
\label{35a}
\end{eqnarray}

\begin{eqnarray}
\tilde{A}^{(0)\mu}_{jik} & = & \frac{\left(z_{i}^{\mu} - z_{j}^{\mu}\right)\zeta_{j}}{\gamma_{ji}^{2}}
\left(a_{0} \xi_{ji}\xi_{ik}\xi_{kj} + b_{0} \zeta_{j}\zeta_{i}\zeta_{k} + c_{0} \xi_{kj}^{2}\zeta_{i}\right)\nonumber\\
& & + \frac{\left(z_{i}^{\mu} - z_{k}^{\mu}\right)\zeta_{k}}{\gamma_{ki}^{2}}
\left(a_{0} \xi_{ji}\xi_{ik}\xi_{kj} + b_{0} \zeta_{j}\zeta_{i}\zeta_{k} + c_{0} \xi_{kj}^{2}\zeta_{i}\right)\nonumber\\
& & + 2 \dot{z}_{i}^{\mu} \left(\zeta_{j}\frac{\gamma_{ij}}{\gamma_{ji}^{2}} + \zeta_{k}\frac{\gamma_{ik}}{\gamma_{ki}^{2}}
+ \frac{\xi_{ij}}{\gamma_{ji}} + \frac{\xi_{ik}}{\gamma_{ki}}\right)\left(\zeta_{j}\zeta_{k} b_{0} + \xi_{kj}^{2} c_{0}\right)\nonumber\\
& & + \dot{z}_{j}^{\mu} \left(\left(\zeta_{j}\frac{\gamma_{ij}}{\gamma_{ji}^{2}} + \zeta_{k}\frac{\gamma_{ik}}{\gamma_{ki}^{2}}
+ \frac{\xi_{ik}}{\gamma_{ki}}\right)\xi_{kj}\xi_{ik} a_{0} 
- \frac{1}{\gamma_{ji}}\left(b_{0} \zeta_{j}\zeta_{i}\zeta_{k} + c_{0} \xi_{kj}^{2}\zeta_{i}\right)\right)\nonumber\\
& & + \dot{z}_{k}^{\mu} \left(\left(\zeta_{j}\frac{\gamma_{ij}}{\gamma_{ji}^{2}} + \zeta_{k}\frac{\gamma_{ik}}{\gamma_{ki}^{2}}
+ \frac{\xi_{ij}}{\gamma_{ji}}\right)\xi_{kj}\xi_{ji} a_{0} 
- \frac{1}{\gamma_{ki}}\left(b_{0} \zeta_{j}\zeta_{i}\zeta_{k} + c_{0} \xi_{kj}^{2}\zeta_{i}\right)\right)
\label{35aa}
\end{eqnarray}

From (\ref{31}, \ref{32}) and (\ref{10}) it follows that, at the second Post-Minkowskian approximation, the metric tensor associated to the particle
$i$ (with non-negligible mass $m_{i}$) is given by the formula:

\begin{eqnarray}
&g^{(i)}_{\mu\nu}& =  \eta_{\mu\nu} - \frac{2G}{c^{2}} \sum\limits^{}_{j \neq i} m_{j} \int d\lambda_{j} \delta(\rho_{ij}) 
\left[\alpha_{0}\dot{z}_{j\mu}\dot{z}_{j\nu} + \beta_{0}\zeta_{j}\eta_{\mu\nu}\right]\nonumber\\
& & - \frac{G^{2}}{c^{4}} \sum\limits^{}_{j \neq i} m_{j}^{2} \int d\lambda_{j} \frac{\delta(\rho_{ij})}{|\eta_{ji}|}
\left[\alpha_{1}\dot{z}_{j\mu}\dot{z}_{j\nu} + \beta_{1}\zeta_{j}\eta_{\mu\nu}\right]\nonumber\\
& & - \frac{G^{2}m_{i}}{c^{4}} \sum\limits^{}_{j \neq i} m_{j} \int d\lambda_{j} \frac{\delta(\rho_{ij})}{|\eta_{ij}|}
\left[\alpha_{1}\dot{z}_{j\mu}\dot{z}_{j\nu} + \beta_{1}\zeta_{j}\eta_{\mu\nu} 
+ \left(\frac{\dot{z}_{i\mu}}{\zeta_{i}} + \frac{(z_{i\mu} - z_{j\mu})}{\gamma_{ij}}\right)
\left(\alpha_{1}\xi_{ij}\dot{z}_{j\nu} + \beta_{1}\zeta_{j}\dot{z}_{i\nu}\right)
\right.\nonumber\\
& & \left.
+ \left(\frac{\dot{z}_{i\nu}}{\zeta_{i}} + \frac{(z_{i\nu} - z_{j\nu})}{\gamma_{ij}}\right)
\left(\alpha_{1}\xi_{ij}\dot{z}_{j\mu} + \beta_{1}\zeta_{j}\dot{z}_{i\mu}\right)
+ \frac{1}{2}\left[\left(\frac{\dot{z}_{i\mu}}{\zeta_{i}} + \frac{(z_{i\mu} - z_{j\mu})}{\gamma_{ij}}\right)
\left(\frac{\dot{z}_{i\nu}}{\zeta_{i}} 
+ \frac{(z_{i\nu} - z_{j\nu})}{\gamma_{ij}}\right) \right.\right.\nonumber\\
& & \left.\left. + \frac{1}{\zeta_{i}}
\left(\eta_{\mu\nu} - \frac{2\dot{z}_{i\mu}\dot{z}_{i\nu}}{\zeta_{i}} 
+ \frac{(z_{i\mu} - z_{j\mu})(z_{i\nu} - z_{j\nu})}{\eta_{ij}^{2}}\right)\right]
\left(\alpha_{1}\xi_{ij}^{2} + \beta_{1}\zeta_{ij}^{2}\right)
\right]
\nonumber\\
& &  - \frac{G^{2}}{c^{4}} \sum\limits^{}_{j \neq i} \sum\limits^{}_{k \neq i, j} m_{j} m_{k}
\int \int d\lambda_{j} d\lambda_{k} \nonumber\\
& & \left[\delta(\rho_{ij}) \delta(\rho_{jk})
\left(\frac{1}{2}a_{0}\xi_{jk}\left(\dot{z}_{j\mu}\dot{z}_{k\nu} + \dot{z}_{k\mu}\dot{z}_{j\nu}\right) 
+ b_{0}\zeta_{j}\zeta_{k}\eta_{\mu\nu}
+ c_{0}\zeta_{j}\dot{z}_{k\mu}\dot{z}_{k\nu}\right) \right.\nonumber\\
& & \left. + \delta(\rho_{jk}) \delta(\rho_{ki}) \left(\frac{1}{2}a_{0}\xi_{jk}\left(\dot{z}_{k\mu}\dot{z}_{j\nu} + \dot{z}_{j\mu}\dot{z}_{k\nu}\right)
+ b_{0}\zeta_{j}\zeta_{k}\eta_{\mu\nu} + c_{0}\zeta_{k}\dot{z}_{j\mu}\dot{z}_{j\nu}\right)
\right. \nonumber\\
& & \left. + \delta(\rho_{ki}) \delta(\rho_{ij}) \left(\frac{1}{2}a_{0}\xi_{jk}\left(\dot{z}_{k\mu}\dot{z}_{j\nu} + \dot{z}_{j\mu}\dot{z}_{k\nu}\right)
+ \left(b_{0}\zeta_{k}\zeta_{j} + c_{0}\xi_{jk}^{2}\right)\eta_{\mu\nu}\right)\right]
\label{10a}
\end{eqnarray}

For the case of a test particle (the mass of which can be neglected) in the presence of $N$ particles with non-negligible masses $m_{i}$ ($i = 1,...,N$)
the above expression simplifies to the following:

\begin{eqnarray}
&g_{\mu\nu}& =  \eta_{\mu\nu} - \frac{2G}{c^{2}} \sum\limits^{}_{i} m_{i}\int d\lambda_{i}\delta(\left(z - z_{i}\right)^{2}) 
\left[
\alpha_{0}\dot{z}_{i\mu}\dot{z}_{i\nu} + \beta_{0}\dot{z}_{i}^{2}\eta_{\mu\nu}\right]\nonumber\\
& & - \frac{G^{2}}{c^{4}} \sum\limits^{}_{i} m_{i}^{2} \int d\lambda_{i} 
\frac{\delta(\left(z - z_{i}\right)^{2})(\dot{z}_{i}^{2})^{\frac{1}{2}}}{|\left(\dot{z}_{i}(z - z_{i})\right)|}
\left[\alpha_{1}\dot{z}_{i\mu}\dot{z}_{i\nu} + \beta_{1}\dot{z}_{i}^{2}\eta_{\mu\nu}\right]\nonumber\\
& &  - \frac{G^{2}}{c^{4}} \sum\limits^{}_{i} \sum\limits^{}_{j \neq i} m_{i} m_{j}
\int \int d\lambda_{i} d\lambda_{j} \nonumber\\
& & \left[
\delta(\left(z - z_{i}\right)^{2}) \delta(\left(z_{i} - z_{j}\right)^{2})
\left(\frac{1}{2}a_{0}\left(\dot{z}_{i}\dot{z}_{j}\right)\left(\dot{z}_{i\mu}\dot{z}_{j\nu} 
+ \dot{z}_{j\mu}\dot{z}_{i\nu}\right)
+ b_{0}\dot{z}_{i}^{2}\dot{z}_{j}^{2}\eta_{\mu\nu} 
+ c_{0}\dot{z}_{i}^{2}\dot{z}_{j\mu}\dot{z}_{j\nu}\right) \right.\nonumber\\
& & \left.
+ \delta(\left(z_{i} - z_{j}\right)^{2}) \delta(\left(z - z_{j}\right)^{2})
\left(\frac{1}{2}a_{0}\left(\dot{z}_{i}\dot{z}_{j}\right)\left(\dot{z}_{i\mu}\dot{z}_{j\nu} 
+ \dot{z}_{j\mu}\dot{z}_{i\nu}\right)
+ b_{0}\dot{z}_{i}^{2}\dot{z}_{j}^{2}\eta_{\mu\nu} 
+ c_{0}\dot{z}_{j}^{2}\dot{z}_{i\mu}\dot{z}_{i\nu}\right)
\right. \nonumber\\
& & \left. 
+ \delta(\left(z - z_{j}\right)^{2}) \delta(\left(z - z_{i}\right)^{2})
\left(\frac{1}{2}a_{0}\left(\dot{z}_{i}\dot{z}_{j}\right)\left(\dot{z}_{i\mu}\dot{z}_{j\nu} 
+ \dot{z}_{j\mu}\dot{z}_{i\nu}\right)
+ \left(b_{0}\dot{z}_{i}^{2}\dot{z}_{j}^{2} 
+ c_{0}\left(\dot{z}_{i}\dot{z}_{j}\right)^{2}\right)\eta_{\mu\nu}\right)
\right]\nonumber\\
\label{10aa}
\end{eqnarray}

{\it The first Post-Newtonian approximation}

The equations of motion (\ref{33}) involve multiple times. The force acting on mass $i$ depends on the state of motion of
particle $i$ at time $t$ and, to account for the time needed for the
transmission of the
interactions, on the states of motion of the remaining $N-1$ particles at
the past and future times $t^{(i,s)}_{j}  (j\neq i, s=-,+)$ and also on $t^{(j,s)}_{k} (k\neq i,j, s=-,+)$.

Using Taylor series expansions involving
the particles' present motions at time $t$, one can rewrite the equations (\ref{33}) using just the one time 
variable $t$\cite{pla}. 
We use series expansions up to terms of second order ($\frac{v^{2}}{c^{2}}$) (first Post-Newtonian approximation (1PN)).

From the definition (\ref{2d}) it follows that:

\begin{equation}
d\lambda_{i} = \frac{c dt \left(1 - \frac{v_{i}^{2}}{c^{2}}\right)^{\frac{1}{2}}}{\zeta_{i}^{\frac{1}{2}}}
\label{C1}
\end{equation}

\begin{equation}
d\lambda_{j} = \frac{c dt_{j} \left(1 - \frac{v_{j}^{2}(t_{j})}{c^{2}}\right)^{\frac{1}{2}}}{\zeta_{j}^{\frac{1}{2}}}
\label{C2}
\end{equation}

The Dirac delta function can be expressed as follows \cite{barut}:

\begin{equation}
\delta\left(c^{2}(t - t_{j})^{2} - (\vec{r}_{i} - \vec{r}_{j})^{2}\right)
= \frac{1}{2 c} \left(
\frac{\delta(t_{j} - t^{(i, -)}_{j})}
{\left(R^{ret}_{ij}
- \frac{\left(\vec{R}^{ret}_{ij} \vec{v}^{(i,-)}_{j}\right)}{c}\right)}
+
\frac{\delta(t_{j} - t^{(i, +)}_{j})}
{\left(R^{adv}_{ij}
+ \frac{\left(\vec{R}^{adv}_{ij} \vec{v}^{(i,+)}_{j}\right)}{c}\right)}
\right)
\label{P17}
\end{equation}

In (\ref{P17}), $t^{(i, s)}_{j}$ ($s = -, +$) are the two roots
of the equation:

\begin{equation}
c^{2}(t - t_{j})^{2} - (\vec{r}_{i}(t) - \vec{r}_{j}(t_{j}))^{2} = 0
\label{P18}
\end{equation}

\noindent and,

\begin{equation}
R^{ret}_{ij} = c \left(t - t^{(i,-)}_{j}\right)
\label{P19}
\end{equation}

\begin{equation}
R^{adv}_{ij} = c \left(t^{(i,+)}_{j} - t\right)
\label{P20}
\end{equation}

\begin{equation}
\vec{R}^{ret}_{ij} = \vec{r}_{i} - \vec{r}^{(i,-)}_{j} 
\label{P21}
\end{equation}

\begin{equation}
\vec{R}^{adv}_{ij} = \vec{r}_{i} - \vec{r}^{(i,+)}_{j}
\label{P22}
\end{equation}

$t - t^{(i,-)}_{j}$ is the time it takes for a signal to travel forward in
time at the speed of light from particle $j$ to particle $i$.

$t^{(i,+)}_{j} - t$ is the time it takes for a signal to travel backward in
time at the speed of light from particle $j$ to particle $i$.

To terms of second order we can write:

\begin{equation}
\vec{r}_{i} - \vec{r}^{(i,-)}_{j} \approx \vec{r}_{ij} + \vec{v}_{j}\frac{r_{ij}}{c} + \vec{v}_{j}\frac{(\vec{r}_{ij}\vec{v}_{j})}{c^{2}}
- \vec{a}_{j}\frac{r_{ij}^{2}}{2 c^{2}},
\label{36a}
\end{equation}

\begin{equation}
\vec{r}_{i} - \vec{r}^{(i,+)}_{j} \approx \vec{r}_{ij} - \vec{v}_{j}\frac{r_{ij}}{c} + \vec{v}_{j}\frac{(\vec{r}_{ij}\vec{v}_{j})}{c^{2}}
- \vec{a}_{j}\frac{r_{ij}^{2}}{2 c^{2}},
\label{36b}
\end{equation}

\begin{equation}
\vec{v}^{(i,-)}_{j} \approx \vec{v}_{j} - \vec{a}_{j}\frac{r_{ij}}{c},
\label{36c}
\end{equation}

\begin{equation}
\vec{v}^{(i,+)}_{j} \approx \vec{v}_{j} + \vec{a}_{j}\frac{r_{ij}}{c}.
\label{36d}
\end{equation}

From (\ref{P18}-\ref{36d}), we find (to terms of second order):

\begin{equation}
\frac{\left(1 - \frac{v_{j}^{(i,-) 2}}{c^{2}}\right)^{\frac{1}{2}}}{\left(R^{ret}_{ij}
- \frac{\left(\vec{R}^{ret}_{ij} \vec{v}^{(i,-)}_{j}\right)}{c}\right)}
\approx \frac{1}{r_{ij}} \left(1 - \frac{(\vec{n}_{ij}\vec{v}_{j})^{2}}{2 c^{2}} - \frac{(\vec{r}_{ij}\vec{a}_{j})}{2 c^{2}}\right)
\label{36e}
\end{equation}

\begin{equation}
\frac{\left(1 - \frac{v_{j}^{(i,+) 2}}{c^{2}}\right)^{\frac{1}{2}}}{\left(R^{adv}_{ij}
+ \frac{\left(\vec{R}^{adv}_{ij} \vec{v}^{(i,+)}_{j}\right)}{c}\right)}
\approx \frac{1}{r_{ij}} \left(1 - \frac{(\vec{n}_{ij}\vec{v}_{j})^{2}}{2 c^{2}} - \frac{(\vec{r}_{ij}\vec{a}_{j})}{2 c^{2}}\right)
\label{36f}
\end{equation}

From the definitions (\ref{2b}, \ref{2c}, \ref{2d}) and (\ref{17}, \ref{18}), it is not difficult to see that, to terms of second order,
we can write:

\begin{equation}
\frac{\xi_{ij}}{\zeta_{ij}} \approx 1 + \frac{v_{i}^{2}}{2 c^{2}} + \frac{v_{j}^{2}}{2 c^{2}} - \frac{(\vec{v}_{i}\vec{v}_{j})}{c^{2}},
\label{37a}
\end{equation}

\begin{equation}
\eta_{ji}^{(i,-)} \approx r_{ij} \left(1 + \frac{(\vec{n}_{ij}\vec{v}_{j})^{2}}{2 c^{2}} + \frac{(\vec{r}_{ij}\vec{a}_{j})}{2 c^{2}}\right),
\label{37b}
\end{equation}

\begin{equation}
\eta_{ji}^{(i,+)} \approx - r_{ij} \left(1 + \frac{(\vec{n}_{ij}\vec{v}_{j})^{2}}{2 c^{2}} + \frac{(\vec{r}_{ij}\vec{a}_{j})}{2 c^{2}}\right),
\label{37c}
\end{equation}

\begin{equation}
\eta_{ij}^{(i,-)} \approx - r_{ij} \left(1 - \frac{(\vec{n}_{ij}\vec{v}_{i})}{c} + \frac{(\vec{n}_{ij}\vec{v}_{j})}{c} + \frac{(\vec{v}_{i} - \vec{v}_{j})^{2}}{2 c^{2}}
+ \frac{(\vec{n}_{ij}\vec{v}_{j})^{2}}{2 c^{2}} - \frac{(\vec{r}_{ij}\vec{a}_{j})}{2 c^{2}}\right),
\label{37d}
\end{equation}

\begin{equation}
\eta_{ij}^{(i,+)} \approx  r_{ij} \left(1 + \frac{(\vec{n}_{ij}\vec{v}_{i})}{c} - \frac{(\vec{n}_{ij}\vec{v}_{j})}{c} + \frac{(\vec{v}_{i} - \vec{v}_{j})^{2}}{2 c^{2}}
+ \frac{(\vec{n}_{ij}\vec{v}_{j})^{2}}{2 c^{2}} - \frac{(\vec{r}_{ij}\vec{a}_{j})}{2 c^{2}}\right).
\label{37e}
\end{equation}

In (\ref{36a} - \ref{37e}), $\vec{r}_{ij} = \vec{r}_{i} - \vec{r}_{j}$ is the relative position of
particle $i$ with respect to particle $j$, $\vec{n}_{ij} \equiv \frac{\vec{r}_{ij}}{r_{ij}}$, $\vec{v}_{i}$ is the velocity
of particle $i$, and $\vec{v}_{j}$, $\vec{a}_{j}$ the velocity and the acceleration of particle $j$. All these quantities are given at time $t$.

From (\ref{8a}) and (\ref{31}), at the first Post-Minkowskian order , for the solutions of the equations of motion we obtain:

\begin{equation}
\zeta_{i} = 1 + \frac{2G}{c^{2}} \sum\limits^{}_{j \neq i} m_{j} \int d\lambda_{j} \delta(\rho_{ij}) \zeta_{j} 
\left(\alpha_{0}\frac{\xi_{ij}^{2}}{\zeta_{ij}^{2}} + \beta_{0}\right) 
\label{38}
\end{equation}

Now, substituting (\ref{C2},\ref{P17}) and (\ref{36e},\ref{36f},\ref{37a}) into (\ref{38}), to terms of second order, 
for the solutions of the equations of motion we can write:

\begin{equation}
\zeta_{i} \approx 1 + \frac{2G}{c^{2}}\left(\alpha_{0} + \beta_{0}\right) \sum\limits^{}_{j \neq i} \frac{m_{j}}{r_{ij}}.
\label{38a}
\end{equation}

Substituting (\ref{C1} - \ref{37e}) and (\ref{38a}) into (\ref{33} - \ref{35aa}), we find the equations of motion to
terms of second order (in $\frac{v^{2}}{c^{2}}$) (first Post-Newtonian approximation):

\begin{eqnarray}
&\vec{a}_{i}& +  G \left(\alpha_{0} + \beta_{0}\right) \sum\limits^{}_{j \neq i} \frac{m_{j}}{r_{ij}^{2}}\vec{n}_{ij}
+ \frac{v_{i}^{2}}{c^{2}} \vec{a}_{i} + \frac{(\vec{v}_{i}\vec{a}_{i})}{c^{2}} \vec{v}_{i} 
+ \frac{2G\alpha_{0}}{c^{2}}\vec{a}_{i} \sum\limits^{}_{j \neq i} \frac{m_{j}}{r_{ij}}
- \frac{G}{2c^{2}}\left(3\alpha_{0} - \beta_{0}\right) \sum\limits^{}_{j \neq i} \frac{m_{j}}{r_{ij}}\vec{a}_{j}
\nonumber\\
& - & \frac{G}{2c^{2}}\left(\alpha_{0} + \beta_{0}\right) \sum\limits^{}_{j \neq i} 
\frac{m_{j}}{r_{ij}}\vec{n}_{ij}(\vec{n}_{ij}\vec{a}_{j})
+ \frac{G}{c^{2}} \sum\limits^{}_{j \neq i} \frac{m_{j}}{r_{ij}^{2}}\vec{n}_{ij}
\left(\alpha_{0} \left(v_{i}^{2} + v_{j}^{2} - 2 (\vec{v}_{i}\vec{v}_{j})\right)
- \frac{3}{2}\left(\alpha_{0} + \beta_{0}\right)(\vec{n}_{ij}\vec{v}_{j})^{2}\right)\nonumber\\
& + & \frac{G}{c^{2}} \left(\beta_{0} - \alpha_{0}\right)\vec{v}_{i}
\sum\limits^{}_{j \neq i} \frac{m_{j}}{r_{ij}^{2}}
\left((\vec{n}_{ij}\vec{v}_{i}) - (\vec{n}_{ij}\vec{v}_{j})\right)
+ \frac{G}{c^{2}} \sum\limits^{}_{j \neq i} \frac{m_{j}}{r_{ij}^{2}}
\vec{v}_{j} \left(2\alpha_{0} (\vec{n}_{ij}\vec{v}_{i}) - \left(\alpha_{0} - \beta_{0}\right) (\vec{n}_{ij}\vec{v}_{j})
\right)\nonumber\\
& + & \frac{G^{2}m_{i}}{c^{2}} \left(\left(\alpha_{0} + \beta_{0}\right)^{2} + \alpha_{1} + \beta_{1}\right)
\sum\limits^{}_{j \neq i} \frac{m_{j}}{r_{ij}^{3}} \vec{n}_{ij}
+ \frac{G^{2}}{c^{2}} \left(2\left(\alpha_{0} + \beta_{0}\right)^{2} + \alpha_{1} + \beta_{1}\right)
\sum\limits^{}_{j \neq i} \frac{m_{j}^{2}}{r_{ij}^{3}} \vec{n}_{ij}\nonumber\\
& + & \frac{G^{2}}{c^{2}} \sum\limits^{}_{j \neq i} \sum\limits^{}_{k \neq i, j} \frac{m_{j} m_{k}}{r_{ij}^{2}} \vec{n}_{ij}
\left(\frac{1}{r_{ik}} \left(2\left(\alpha_{0} + \beta_{0}\right)^{2} + a_{0} + b_{0} + c_{0}\right)
+ \frac{1}{r_{jk}} \left(\left(\alpha_{0} + \beta_{0}\right)^{2} + a_{0} + b_{0} + c_{0}\right)
\right)\nonumber\\
& = & 0.
\label{39}
\end{eqnarray}

Complete agreement with the equations of motion of General Relativity \cite{einstein, landau}, 
at the first Post-Newtonian order, is achieved if:

\begin{equation}
\alpha_{0} = 2,
\label{40a}
\end{equation}

\begin{equation}
\beta_{0} = -1,
\label{40b}
\end{equation}

\begin{equation}
\alpha_{1} + \beta_{1} = -2,
\label{40c}
\end{equation}

\begin{equation}
a_{0} + b_{0} + c_{0} = -2.
\label{40d}
\end{equation}

\vskip 2cm

{\it Conclusions}

We have obtained Lorentz invariant equations of motion describing the gravitational interactions
of a system consisting of $N$ point masses.
The equations are derived explicitly from a Lorentz invariant action.
Contrary to General Relativity, which is a field theory, the model presented 
here is a relativistic action-at-a-distance description (the interactions are not mediated by a field).
We have shown that the equations of motion for $N$ point masses 
agree with those of General Relativity at the first Post-Newtonian approximation (1PN). 
Agreement with General Relativity for the $N$ body problem at orders beyond 1.5PN
has not been established. 
The model presented is in agreement with General Relativity for the one-body case, 
at all orders. At the first Post-Minkowskian approximation our model reduces to the model of Havas and Golberg
\cite{golberg, havas}, which is known to be in agreement with General Relativity in this approximation.
Due to this agreement, gravitational radiation effects in our model begin to appear at the 2.5PN order
($\frac{v^{5}}{c^{5}}$) \cite{smith, walker}.

\end{document}